\documentclass[iop,apj,appendixfloats]{emulateapj}
\usepackage{apjfonts}

\gdef\msun{$M_{\odot}$}
\lefthead{van Dokkum et al.}
\righthead{}
\slugcomment{Accepted for publication in ApJ Letters}
\begin{document}

\title{The Assembly of Milky Way-Like Galaxies Since $z\sim 2.5$}

\author{Pieter G.\ van Dokkum\altaffilmark{1},
Joel Leja\altaffilmark{1},
Erica June Nelson\altaffilmark{1},
Shannon Patel\altaffilmark{2},
Rosalind E.\ Skelton\altaffilmark{1},
Ivelina Momcheva\altaffilmark{1},
Gabriel Brammer\altaffilmark{3},
Katherine E.\ Whitaker\altaffilmark{4},
Britt Lundgren\altaffilmark{5},
Mattia Fumagalli\altaffilmark{2},
Charlie Conroy\altaffilmark{6},
Natascha F\"orster Schreiber\altaffilmark{7},
Marijn Franx\altaffilmark{2},
Mariska Kriek\altaffilmark{8},
Ivo Labb\'e\altaffilmark{2},
Danilo Marchesini\altaffilmark{9},
Hans-Walter Rix\altaffilmark{10},
Arjen van der Wel\altaffilmark{10},
Stijn Wuyts\altaffilmark{7}
}

\altaffiltext{1}
{Department of Astronomy, Yale University, New Haven, CT 06511, USA}
\altaffiltext{2}
{Leiden Observatory, Leiden University, Leiden, The Netherlands}
\altaffiltext{3}
{European Southern Observatory, Alonson de C\'ordova 3107, Casilla
19001, Vitacura, Santiago, Chile}
\altaffiltext{4}
{Astrophysics Science Division, Goddard Space Center, Greenbelt,
MD 20771, USA}
\altaffiltext{5}
{Department of Astronomy, University of Wisconsin-Madison, Madison,
WI 53706, USA}
\altaffiltext{6}
{Department of Astronomy \& Astrophysics, University of California,
Santa Cruz, CA, USA}
\altaffiltext{7}
{Max-Planck-Institut f\"ur extraterrestrische Physik, Giessenbachstrasse,
D-85748 Garching, Germany}
\altaffiltext{8}
{Department of Astronomy, University of California, Berkeley, CA 94720, USA}
\altaffiltext{9}
{Department of Physics and Astronomy, Tufts University, Medford, MA 02155,
USA}
\altaffiltext{10}
{Max Planck Institute for Astronomy (MPIA), K\"onigstuhl 17, D-69117,
Heidelberg, Germany}

\begin{abstract}

Galaxies with the mass of the Milky Way
dominate the stellar mass density of the Universe but it is
uncertain how and when they were assembled. Here we study 
progenitors of these galaxies  out to $z=2.5$,
using data from the 3D-HST and CANDELS Treasury surveys. We find that
galaxies with present-day stellar masses of $\log (M)\approx 10.7$
built $\sim 90$\,\% of their stellar
mass since $z=2.5$, with most of the star formation
occurring before $z=1$. 
In marked contrast to the assembly history of
massive elliptical galaxies, mass growth is not limited to large radii:
the mass in the central 2\,kpc of the galaxies increased by a factor
of $3.2^{+0.8}_{-0.7}$ between $z=2.5$ and $z=1$. 
We therefore rule out simple models in which bulges were fully assembled
at high redshift and
disks gradually formed around them. Instead, bulges (and black
holes) likely
formed in lockstep with disks, through bar instabilities, migration,
or other processes. We find that after $z=1$ the growth
in the central regions gradually stopped
and the disk continued to build, consistent with recent studies of the gas
distributions in $z\sim 1$
galaxies and the properties of many spiral galaxies today.

\end{abstract}


\section{Introduction}

The Milky Way is a very typical galaxy, in the sense that a randomly
chosen star in the Universe is most often found in a bulge-disk
system of similar mass.
Despite their ubiquity, and our exquisite knowledge of one example
of their class, the assembly history of large spiral
galaxies is still uncertain
(see {Rix} \& {Bovy} 2013, and references therein).
A key question is when  different structural components
of the galaxies were formed.
The morphology and stellar populations of many spiral galaxies suggest a two-phase
scenario, with bulges typically forming at high redshift
and disks gradually assembling around them
(e.g., {Kauffmann}, {White}, \&  {Guiderdoni} 1993;
{Zoccali} {et~al.} 2006).
Such a purely inside-out scenario would be qualitatively
similar to  the assembly history of
massive ellipticals, which formed a dense core at
high redshift and subsequently built up their outer parts
(e.g., {van Dokkum} {et~al.} 2010; {Hilz}, {Naab}, \& {Ostriker} 2013).

However, the structural evolution of spiral galaxies is
probably more complex than this.
In cosmological
simulations of gas accretion
the structure of the forming galaxy
not only depends on the properties of the dark matter
but also on the details of the feedback mechanism
(e.g., {Agertz}, {Teyssier}, \& {Moore} 2011; {Brooks} {et~al.} 2011) and on
the accretion mode (e.g., {Sales} {et~al.} 2012).
Furthermore, major mergers may be too rare to form many
bulges (e.g., {Kitzbichler} \& {White} 2008), and several studies
have suggested alternative ways to 
build up central mass concentrations.
In particular, (pseudo-)bulges may be the result of secular evolution
(e.g., Kormendy \& Kennicutt 2004; Parry et al.\ 2009), ``direct injection''
of gas in cold streams (e.g., {Sales} {et~al.} 2012)
and/or migration
in unstable disks
({Elmegreen}, {Bournaud}, \&  {Elmegreen} 2008; {Dekel} {et~al.} 2009; {Krumholz} \& {Dekel} 2010).
Such clumpy, unstable, rapidly star-forming disks have been shown to
exist at high redshift (e.g., {Genzel} {et~al.} 2008; {F{\"o}rster Schreiber} {et~al.} 2011).

In this paper we provide new constraints on the assembly
of spiral galaxies by studying plausible progenitors of Milky Way-mass
galaxies in the 3D-HST survey ({Brammer} {et~al.} 2012).
The goals
are to determine the average star formation histories of
these galaxies, to determine the mass growth
in their central regions
since $z=2.5$, and to compare their structural evolution to
that of more massive galaxies.
The data also
provide key constraints on the ingredients
in recent hydrodynamical models: these models now succeed
in reproducing many of the properties of the present-day
Milky Way ({Brooks} {et~al.} 2011; {Guedes} {et~al.} 2011) and to
improve them further we need to test their predictions
at earlier times. A {Kroupa} (2001) IMF is assumed throughout
the paper.

\begin{figure*}[htbp]
\epsfxsize=16.3cm
\epsffile[-20 170 520 586]{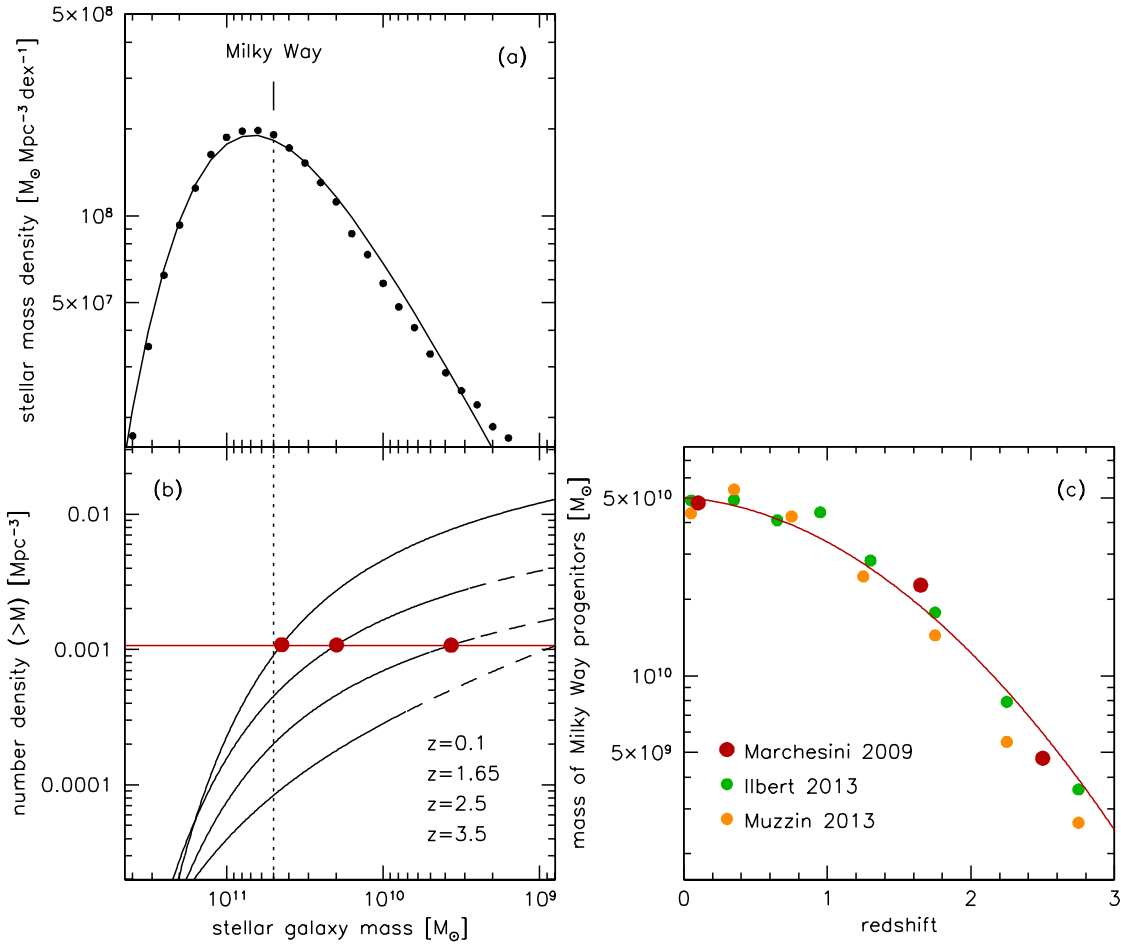}
\caption{\small
{\em (a)} Stellar mass density of the Universe as a function of galaxy
mass, as determined from the SDSS-GALEX $z=0.1$ mass function
of {Moustakas} {et~al.} (2013).
{\em (b)}
Evolution of the cumulative
galaxy mass function from $z=0.1$ to $z=3.5$
(SDSS-GALEX and Marchesini et al.\ 2009).
The horizontal line indicates a constant cumulative co-moving number density
of $1.1\times 10^{-3}$\,Mpc$^{-3}$.
{\em (c)} Mass evolution at a constant number density of
$1.1\times 10^{-3}$\,Mpc$^{-3}$.
\label{sel.fig}}
\end{figure*}

\section{Mass Evolution}

Following previous studies
(e.g., {van Dokkum} {et~al.} 2010; {Papovich} {et~al.} 2011; {Patel} {et~al.} 2013; {Leja}, {van Dokkum}, \& {Franx} 2013)
we link progenitor and descendant galaxies
by requiring that they have the same (cumulative)
co-moving number density.
Effectively, galaxies are ranked according to their stellar mass and we
study galaxies at high redshift that have the same
rank order as the Milky
Way does at $z=0$. The implicit assumption is that rank order is conserved
through cosmic time, or that processes that break the rank order do not
have a strong effect on the average measured properties.
As shown in {Leja} {et~al.} (2013) the method recovers the true mass evolution
of galaxies remarkably well in simulations that include  merging,
quenching, and scatter in the growth rates of galaxies.

The present-day stellar mass of the Milky Way is approximately
$5\times 10^{10}$\,\msun\ ({Flynn} {et~al.} 2006; {McMillan} 2011).
Using the SDSS-GALEX stellar galaxy
mass function of {Moustakas} {et~al.} (2013) we find
that galaxies with masses
$>5\times 10^{10}$\,\msun\ have a number density of 
$1.1\times 10^{-3}$\,Mpc$^{-3}$.
We then trace the progenitors of these  galaxies by identifying,
at each redshift, the mass for which the cumulative number density
is $1.1\times 10^{-3}$\,Mpc$^{-3}$
(see Fig.\ \ref{sel.fig}b). We used the {Marchesini} {et~al.} (2009)
mass functions as they are complete in the relevant mass and
redshift range; we
verified that the results are similar when other mass functions
are used ({Ilbert} {et~al.} 2013; {Muzzin} {et~al.} 2013).

The stellar mass evolution for galaxies with the rank order of the Milky
Way is shown in Fig.\ \ref{sel.fig}c. The evolution is
rapid from $z\sim 2.5$ to $z\sim 1$ and relatively slow afterward.
We therefore approximate the evolution with a quadratic function of the
form
\begin{equation}
\log(M_{\rm MW}) = 10.7 -0.045 z - 0.13 z^2.
\label{massevo.eq}
\end{equation}
Based on the variation between mass functions of different authors,
and the results of {Leja} {et~al.} (2013),
we estimate that the uncertainty in the evolution out to
$z\sim 2.5$ is approximately 0.2 dex.\footnote{We verified that changing
the evolution does not affect the key results of this paper.}
More than half of the present-day
mass  was assembled in the 3\,Gyr
period between $z=2.5$ and $z=1$, and
as we show later the mass growth is likely dominated by star formation
at all redshifts.
The mass evolution is significantly faster than that of more massive galaxies
({van Dokkum} {et~al.} 2010; {Patel} {et~al.} 2013), consistent with recent results
of {Muzzin} {et~al.} (2013).

\begin{figure*}[htbp]
\epsfxsize=16cm
\epsffile[43 185 516 640]{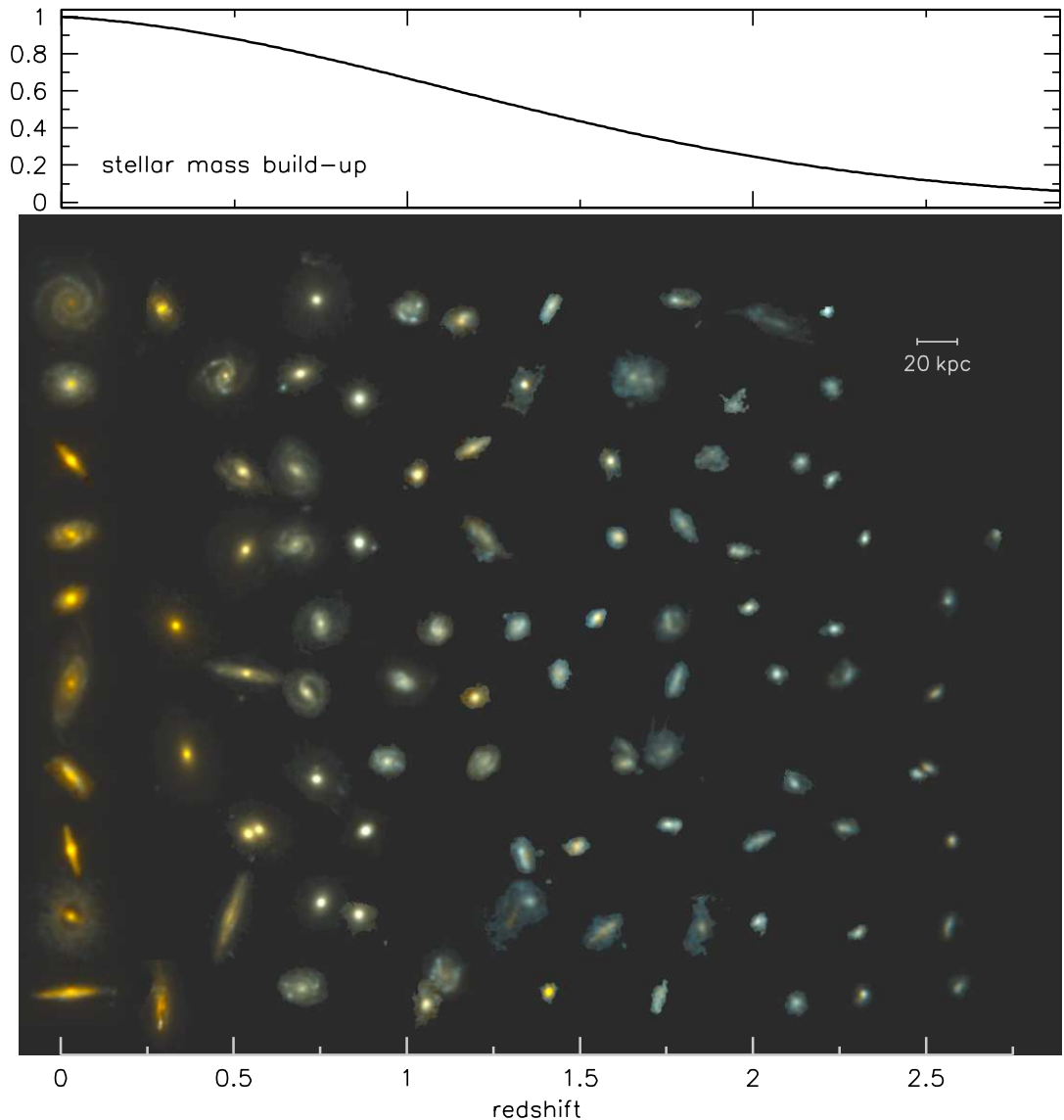}
\caption{\small
Examples
of galaxies with the number density of the Milky Way at $0<z<2.75$.
Galaxies at $z\approx 0.015$ are from the SDSS;
galaxies at higher redshift are from the 3D-HST and CANDELS surveys.
The color images
were created from data in the same
rest-frame bands ($u$ and $g$) at all redshifts and have a common
physical scale. Their intensities are scaled
so they are proportional to
mass, indicated in the top panel. Galaxies at high redshift
have relatively low surface densities; their centers and outer
parts
seem to build up at the same time, at least until $z\sim 1$.
\label{mos.fig}}
\end{figure*}

\section{Milky Way Progenitors from $z=0$ to $z=2.5$}

\subsection{Rest-frame Images}

Having determined the stellar mass evolution with
redshift, we can now select galaxies in mass bins centered on this evolving
mass and study how their properties changed.
We selected galaxies in  GOODS-North and GOODS-South  as these
fields have multi-band ACS and WFC3 imaging
(from the GOODS and CANDELS
surveys respectively; {Giavalisco} {et~al.} 2004; {Grogin} {et~al.} 2011; {Koekemoer} {et~al.} 2011), as
well as WFC3 G141 grism spectra from the 3D-HST program ({Brammer} {et~al.} 2012).
Redshifts, stellar masses, and star formation rates were determined
from deep photometric catalogs in these fields, combined with the
grism spectra (see Brammer et al.\ 2012 and references therein, and
R.\ Skelton et al., in preparation). The 3D-HST
v2.1 catalogs are $\approx 100$\,\%
complete in the relevant mass and redshift range, but we note that we rely
largely on photometric redshifts (rather than grism
redshifts) at $z\gtrsim 1.3$.

There are 361 galaxies at $0.25<z<2.75$ in the catalogs
whose mass is within $\pm 0.1$ dex of
$M_{\rm MW}(z)$. Images of a random subset of 90 
are shown in Fig.\ \ref{mos.fig}. The images have the same
physical scale and represent the same  rest-frame
filters ($u$ and $g$). Their brightness is scaled in such a
way that their total ($u$ + $g$) flux is proportional to
$M_{\rm MW}(z)$. The rest-frame $u$ and $g$
images were created by interpolating the
two ACS and/or WFC3 images (smoothed to the $H_{160}$ resolution)
whose central wavelengths are closest
to the redshifted $u$ and $g$ filters.

Also shown are nearby galaxies from
the Sloan Digital Sky Survey (SDSS). We selected 40 galaxies with
$0.013<z<0.017$ and $10.62 < \log M < 10.78$ from the DR7
MPA-JHU catalogs\footnote{http://home.strw.leidenuniv.nl/\~{ }jarle/SDSS/}
({Brinchmann} {et~al.} 2004), and degraded their $u$ and $g$
images to the same spatial resolution as the high redshift galaxies.
A random subset of 10 are shown in Fig.\ \ref{mos.fig}.

\begin{figure*}[htbp]
\epsfxsize=17cm
\epsffile{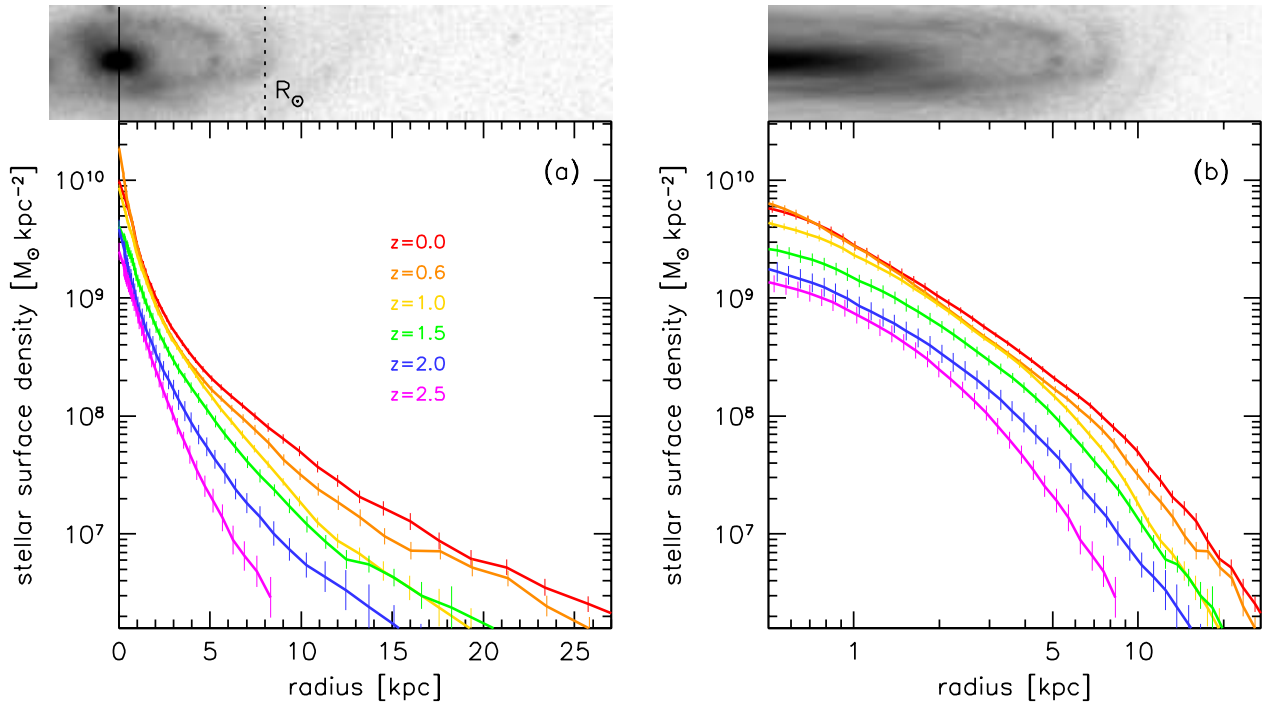}
\caption{\small
Surface density profiles from $z=2.5$ to $z=0$,
as measured from averaged, PSF-corrected rest-frame
$g$ band images in each redshift bin. The horizontal
axis is linear in {\em (a)} and logarithmic in {\em (b)}.
The galaxy image is randomly chosen from our SDSS sample to
illustrate the radial extent of the profiles. The main
evolution is in normalization, which is
determined by $M_{\rm MW}(z)$ (Eq.\ 1).
The profile {\em shapes} are very similar
from $z\sim 2.5$ to $z\sim 1$,
which implies that the galaxies are building up
mass at all radii. After $z\sim 1$
the central regions gradually stop growing but the disk continues to build.
\label{profs.fig}}
\end{figure*}

It is clear from Fig.\ \ref{mos.fig} that present-day galaxies
with the mass of the Milky Way have changed 
over cosmic time. The most obvious change is that
galaxies became redder with time, particularly after
$z\sim 1$, indicative of a decrease in the specific star formation rate.
The galaxies also appear brighter at lower redshift in
Fig.\ \ref{mos.fig}, reflecting the
mass evolution of Eq.\ \ref{massevo.eq}. A striking aspect of this change
in brightness, and a central result of this paper,  is that the
bulges appear to change nearly as much as the disks, particularly
at $z>1$. We do {\em not} see high density ``naked bulges''
at $z\sim 2$ around which disks gradually assembled. Instead,
the central densities at $z\sim 2$ were much lower
than the central densities at $z\sim 0$. We quantify this result
in the remainder of the paper.

\subsection{Evolution of Surface Density Profiles}

We first analyze the surface density profiles of the galaxies,
to study their mass growth as a function of radial distance from
their centers. Following {van Dokkum} {et~al.} (2010) 
we measured the profiles from stacked images to increase the signal-to-noise
ratio. The galaxies were grouped in six bins with mean redshifts $0.015$,
$0.60$, $1.0$, $1.5$, $2.0$, and $2.4$. Each bin contains 40--90
galaxies. The rest-frame $u$ and $g$ band images in each bin
were normalized and stacked, aggressively masking all neighboring objects.

The image stacks were corrected for the effects of the point spread
function (PSF) following the method outlined in {Szomoru} {et~al.} (2010).
First, a two-dimensional {Sersic} (1968) model, convolved with the
PSF, was fit to the stacks using the GALFIT code ({Peng} {et~al.} 2010).
Then the residuals of this fit were added to the {\em un}convolved
Sersic model. As shown in {Szomoru} {et~al.} (2010) this method reconstructs
the true flux distribution with high fidelity,
even for galaxies that are poorly fit by Sersic profiles.
The resulting radial surface density profiles
are shown in Fig.\ \ref{profs.fig}. The profiles are derived
from the rest-frame $g$ band images
and scaled such that the total mass within a diameter
of 50\,kpc is equal to $M_{\rm MW}(z)$. Errorbars were determined from
bootstrapping (see van Dokkum et al.\ 2010). We note here that
the $u-g$ color gradients of the stacks
are small ($\approx 0.1$\,dex$^{-1}$)
at all redshifts, consistent with
other studies (e.g., {Szomoru} {et~al.} 2013).

There is strong evolution in the overall normalization of the
profiles from $z=2.5$ to $z=1$ and less evolution
thereafter, reflecting the mass evolution of Eq.\ \ref{massevo.eq}.
The evolution from $z=2.5$ to $z=1$ is strikingly uniform: the
profiles are roughly parallel to one another in
Fig.\ \ref{profs.fig}b, and rather than 
assembling only inside-out the galaxies increase their mass at all
radii. This is in marked contrast
to more massive galaxies, which form their cores early and
exclusively build up their outer parts over this redshift range
(see Fig.\ 6 in van Dokkum et al.\ 2010 and Fig.\ 6 in Patel
et al.\ 2013). After $z\sim 1$ the evolution
in the central parts slows down
but the outer parts continue to build up, consistent with
the visual impression that around
this time the classical ``quiescent bulge and
star forming disk'' structure of spiral galaxies was established
(see Fig.\ \ref{mos.fig}).

\begin{figure*}[htbp]
\epsfxsize=17cm
\epsffile[20 438 552 672]{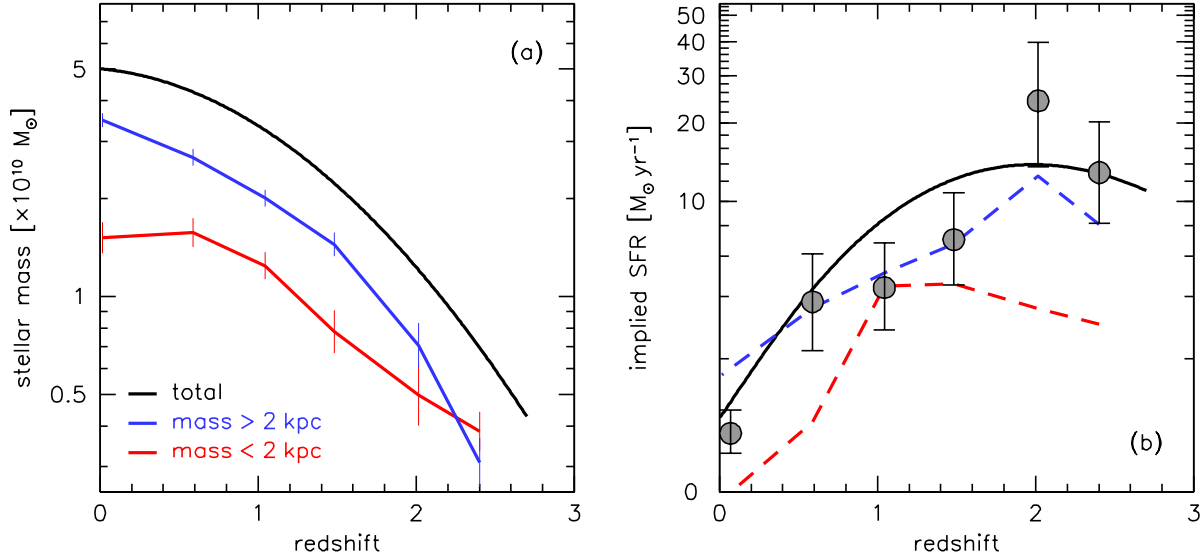}
\caption{\small
{\em (a)} Comparison of the mass growth in the central regions to the
growth at larger radii. The galaxies grow at all radii until $z\sim 1$,
after which the mass inside $r=2$\,kpc remains roughly constant.
{\em (b)} Implied evolution of the star formation rate. Data points
are the mean measured star formation rates of the galaxies in each
redshift bin, from the 3D-HST v2.1 catalogs (Skelton et al., in
preparation). There is an excellent match
between the black curve and the points, indicating that mergers are
not required to explain the mass evolution of large spiral galaxies.
\label{grow.fig}}
\end{figure*}

\subsection{Mass Growth at Different Radii}

We explicitly show the mass growth
at different radii in Fig.\
\ref{grow.fig}a. From $z=2.5$ to $z=1$ the mass outside of $r=2\,$kpc
increased by $0.8 \pm 0.1$ dex and the mass inside 2\,kpc
increased by $0.5 \pm 0.1$ dex. Although the mass evolution is
slightly faster at large radii than at small radii,
the trend is qualitatively different from that seen in
more massive galaxies: after $z\sim 2$ the mass within
2\,kpc is constant to within 0.1\,dex for galaxies
with $\log(M/M_{\rm \odot})(z=0) = 11.2$ (see Fig.\ 7 of Patel et al.\ 2013).
At later times the central mass growth decreases:
from $z=1$ to $z=0$ the mass within 2\,kpc 
grows by only $0.09 \pm 0.04$\,dex.
 
In Fig.\ \ref{grow.fig}b we express the growth in mass as an (implied)
star formation rate. The star formation rate was calculated directly from
Eq.\ \ref{massevo.eq}, with a $\times 1.35$ upwards correction to account
for mass loss in winds.\footnote{This factor is the
mass loss after 2\,Gyr for a Kroupa (2001) IMF.} The implied star
formation rate is approximately constant at $10-15$\,\msun\,yr$^{-1}$
from $z\sim 2.5$ to $z\sim 1$ and then decreases rapidly 
to $\lesssim 2$\,\msun\,yr$^{-1}$ at $z=0$. The form of this
star formation history is well approximated by
\begin{equation}
\log (1+{\rm SFR}) = 0.26 + 0.92 z - 0.23 z^2.
\label{sfr.eq}
\end{equation}
We can compare Eq.\ \ref{sfr.eq}
with the actual star formation rates of
the galaxies: the points with errorbars in Fig.\ \ref{grow.fig}b
show the mean star formation
rates of the galaxies that went into the analysis,
as obtained from SED fits
(see {Kriek} {et~al.} 2009, and Skelton et al., in preparation).
With $\chi^2=7.3$ and 5 degrees of freedom the points are consistent with
the solid line. This consistency is reassuring, and also
implies that the assembly history
can be fully explained by star formation, with
mergers likely playing a minor role. This can, again, be
contrasted with more massive galaxies, as star formation is
not sufficient to explain their growth after $z\sim 1.5$
({van Dokkum} {et~al.} 2010). 

\subsection{Structural Evolution}
\begin{figure*}[htbp]
\epsfxsize=16cm
\epsffile[10 184 540 658]{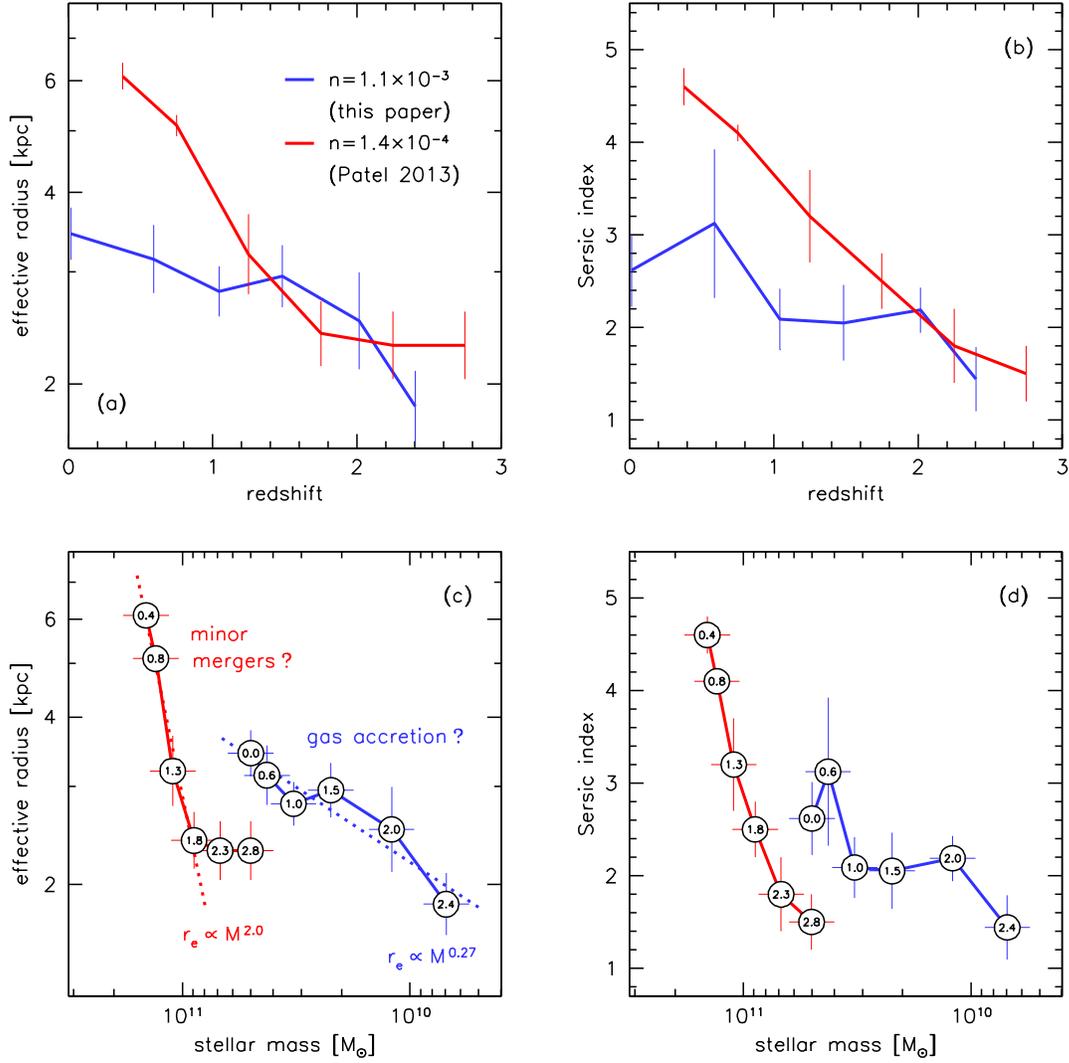}
\caption{\small
Effective radius and Sersic index as a function of redshift and mass,
for Milky Way progenitors (blue) and more massive galaxies (red,
taken from Patel et al.\ 2013). Galaxies like the Milky
Way have
undergone much less structural evolution than the giant elliptical
galaxies that populate the high mass end of the mass function.
\label{structure.fig}}
\end{figure*}

Finally, we quantify the implications of our results for the structural
evolution of galaxies with the present-day
mass of the Milky Way. As
the mass growth is mostly independent of
radius, we expect the structure of the galaxies to remain more
or less the same over cosmic time. The evolution of the GALFIT-derived
structural parameters of the stacks
(see \S\,3.2) is shown in Fig.\ \ref{structure.fig}.

The effective radii and Sersic indices have indeed changed relatively
little since $z\sim 2.5$, particularly when it is considered that the
galaxies increased in mass by a factor of $\sim 10$ over this time.
The radius increased by a factor $\sim 1.8$ and the Sersic index
changed from $n\sim 1.5$ to $n\sim 2.5$. The red curves
show the change in these same parameters for high mass galaxies,
calculated in the same way ({Patel} {et~al.} 2013). Even though the
progenitors of today's
massive galaxies increased their mass by only a factor of $\sim 3$ over
this redshift range they show much more dramatic structural
evolution.

This point is emphasized in Fig.\ \ref{structure.fig}c,d which compares
the structural evolution to the mass evolution for both classes
of galaxies. The sizes of massive galaxies grow as $r_e
\propto M^{2.0\pm 0.1}$
({van Dokkum} {et~al.} 2010; {Hilz} {et~al.} 2013; {Patel} {et~al.} 2013), whereas
those of galaxies with the mass of the
Milky Way grow as $r_e\propto M^{0.27 \pm 0.04}$.
This slope is similar to that of the size-mass relation of
late type galaxies (e.g., {Shen} {et~al.} 2003).
We note that an increase in Sersic index does {\em not}
imply growth of a classical bulge for either class
of galaxy
(see also {Nelson} {et~al.} 2013).

\section{Discussion}

In this {\em Letter} we have demonstrated that it is possible to
obtain a description of the formation 
of galaxies with the mass of the Milky Way all the way
from $z\sim 2.5$ to the present.
We find that these galaxies built up
$\sim 90$\,\% of their stellar mass since $z\sim 2.5$. The build-up
can be fully explained by the measured star formation rates of
the galaxies, and does not require significant merging.
A key result of our paper is that the mass growth took place
in a fairly uniform way, with the galaxies increasing their mass
at all radii. Our results are therefore
inconsistent with simple models in which the central parts
of spiral gaaxies are fully assembled at early times:
we do not find ``naked bulges'' at high redshift. Instead,
they are consistent with models in which bulges (and presumably
black holes)
were largely built up at the same
time as disks, through short-lived peaks in the accretion rate,
bar instabilities,  migration, or other processes
(e.g., {Kormendy \& Kennicutt} 2004; {Dekel} {et~al.} 2009).
The implied star formation rate declines
precipitously after $z\sim 1$, particularly in the central
$\approx 2$\,kpc of the galaxies. By $z=0$
we are left with quiescent bulges and slowly star forming disks.

Many other studies have reached similar conclusions using
independent arguments; here we limit
the discussion to a handful examples.
{Wuyts} {et~al.} (2011) and {Nelson} {et~al.} (2013)  find
that star formation at high redshift typically
occurs in disks.  {Nelson} {et~al.} (2012) find that 
galaxies begin to build inside-out at $z\sim 1$.  As noted in
\S\,1
{Genzel} {et~al.} (2008), {F{\"o}rster Schreiber} {et~al.} (2011), and others have
identified thick, clumpy star forming disks at $z\sim 2$.
Finally, the inferred
star formation history (Eq.\ 2) is broadly
consistent with results from other methods
(e.g., Yang et al.\ 2012; {Behroozi}, {Wechsler}, \&  {Conroy} 2012).

It is tempting to compare our results directly to known properties
of the Milky Way itself; e.g., 
Eq.\ \ref{sfr.eq} implies a $z=0$ star formation rate of
$\sim 1$\,\msun\,yr$^{-1}$, in reasonably good agreement
with that of the Milky Way
({Robitaille} \& {Whitney} 2010). We note, however, that the Milky Way
has a relatively low bulge-to-disk ratio for its mass
(e.g., {McMillan} 2011). Furthermore, the Milky Way,
like any other galaxy,
has had a unique history and it is fundamentally
hazardous to apply the statistical analysis of samples of distant
galaxies to an individual nearby galaxy (see, e.g., Fig.\ 1 of
Leja et al.\ 2013).

As noted in previous Sections, the formation process of galaxies with
$\log M \approx 10.7$
appears to be very different from that of more massive
galaxies. Massive
galaxies formed exclusively
inside-out since $z\sim 2$,
with their extended wings assembling after formation
of a compact core at earlier times.
It will be interesting to see
if galaxy formation models can reproduce both types of behavior
seen in Fig.\ \ref{structure.fig}; e.g., it may be
that (minor) mergers
lead to growth at large radii whereas gas accretion leads to
more uniform growth.

This study can be extended and improved in many ways. Most importantly,
we have largely ignored systematic uncertainties in our analysis.
Among the uncertainties are the low mass end of the mass function
at $z>2$ (see, e.g., {Reddy} \& {Steidel} 2009); possible errors in the
number density selection technique ({Leja} {et~al.} 2013); systematic
errors in redshifts and/or masses in the 3D-HST v2.1 catalogs;
and the conversion
of light-weighted to mass-weighted profiles. We have also ignored
the spread in galaxy properties at fixed mass
(see, e.g., {Baldry} {et~al.} 2006; {Franx} {et~al.} 2008, and Fig.\ 2).
Finally, our analysis
is, by its nature, indirect: we do not actually observe
the formation of different parts of the galaxies but infer this
from changes in their stellar surface densities.
Stellar migration and other processes almost certainly
altered the orbits of stars after their formation ({Ro{\v s}kar} {et~al.} 2008).
Deep, direct observations
of spatially-resolved gas
distributions at high redshift, particulary in the crucial
epoch $1<z<2.5$, are needed to disentangle formation and migration,
and to shed light on the physical processes
that are at work (e.g., {Nelson} {et~al.} 2012, 2013; {Freundlich} {et~al.} 2013).

\begin{acknowledgements}
We thank the referee for an excellent report which improved the paper.
Support from STScI grant GO-1277 is gratefully acknowledged.
\end{acknowledgements}


\end{document}